# Scams in modern societies: how does China differ from the world?

Jeff Yan

IDA, Linköping University, Sweden
`jeff.yan@liu.se`

**Abstract.** We study a set of high-profile scams that were well-engineered and have hit people hard in China in recent years. We propose a simple but novel theoretical framework to examine *psychological*, *situational* and *social fabric* factors that have played a role in these scams. We also use this framework as a tool to explore scam countermeasures. In so doing, we identify how these Chinese scams differ from their Western counterparts.

**Keywords:** Deception, social engineering, mind manipulation, cross-cultural studies, banking security, socio-technical security

## 1   Introduction

Various modern scams have been rampant in China in the recent years, with victims incurred serious financial loss and undergone psychological trauma. Many single cases even made national headlines.

It is striking that some of these scams appear to have never occurred in the western countries; at least no reported incident or analysis is known to us. They also appear to constitute novel genres different from those discussed in the literature.

In this paper, we introduce a selected set of well-engineered, high-profile scams in China. Our selection criteria include popularity, visibility and representativeness, judged by the media coverage, the scale of scam, the amount of financial loss caused (as reported), and the volume of discussions in online forums.

For each scam, we present a brief but anatomic analysis, including: 1) What has happened? We provide a brief summary of scam setup, procedure and outcome; and 2) Why has the scam worked?

We propose a novel three-dimensional framework to explain why these scams have worked. In our analysis, mind manipulation tricks on individual victims, situational factors, and weaknesses in the social fabrics (including infrastructure in the society) have all played a role to the advantage of the scammers. These three dimensions do not necessarily work alone. Instead, they interact with each other, and often the last two dimensions appear to reinforce the first dimension, creating a strong psychological effect to trap people into victims.

Moreover, we apply our framework to explore countermeasures for addressing the scams. We also attempt to identify what makes unique features in the Chinese scams, and what makes these scams less or not effective in the West.

Our study is perhaps the first account of high-profile modern scams in China to an international audience, and the first academic analysis of these scams. This study is also the first step towards a comparative study of the scam landscape in contemporary China vs. in the western countries.

## 2 High-profile scams in China: new genres

We introduce three genres of modern scams in China, namely the criminal investigating scam, the flight rebook scam and the cab sharing scam. Each has made headlines of national news, TV channels or major websites, e.g. covered by major media outlets such as Xinhua News Agency (the biggest and most influential media organization in China), China Central Television (commonly abbreviated as CCTV, the predominant state television broadcaster in China), and websites such as Sina.com and QQ.com (each of which has a reader base of hundreds of millions).

Each of the scam genres has reportedly caused the victims to lose millions of RMB (1 US$ = 6 RMB) or more, with the financial loss incurred by the criminal investigation scam being the most serious, reaching an estimate of hundreds of millions RMB.

Google search returns about 2,710,000 hits for the criminal investigating scam (with simple keywords "警察来电"), about 439,000 hits for the cab sharing scam (with simple keywords "拼车诈骗") and about 477,000 hits for the flight rebook scam (with simple keywords: "机票改签 诈骗"). For convenience, we only use a single keyword for each scam but ignore many of its derived variant names. We believe many more hits will be returned by Google, if our search also covers some of those variants.

### 2.1 Criminal investigation scam

Many variants of this scam have been reported. A typical account goes as follows.

A scammer phones a victim, impersonating a bank employee: "Your account is used for money laundering (or bribery, or corruption, or other accusations) and it's now under custody and criminal investigation".

The victim denies wrongdoing.

The scammer pushes back, "Well, it could be your wrongdoing, or somebody else had stolen your account and misused it".

The scammer continues, "I can now ring for you the very police officer who is investigating your case."

Then, the victim hears the scammer making a phone call, and then hears a standard greeting message "this is X police station, and your call will be connected shortly".

The victim's phone shows that the current call is connected from that police station's number.

A police officer makes a self-introduction, saying that the victim's case is serious and can lead to ten years of imprisonment;

Shortly, a URL is sent to victim, linking to an arrest warrant issued to the victim's name and address, stamped by state agencies.

The victim is then told: to avoid an imminent arrest, you need transfer a large sum to a designated account as soon as possible.

Often, the victim complies.

In some cases, each victim lost an astonishing sum, up to millions of US$ [1]. Highly-educated people have fallen into victims, too.

**2.2   Cab sharing scam**

Cab sharing has been a common practice in some parts of China, not just for personal cost saving, but for the social benefit of reducing pollution and traffic jam, as encouraged by the government. Scammers found themselves a niche in cab sharing, too. A high-profile scam had hit hard a Chinese city. In two months (May--June 2009), nearly 200 incidents were reported to police in this city, amounting to a loss of over 5 million RMB (about US$800K). The police reportedly estimated that about 1000 people fell into victims in this particular city. The victims were all female.

The scam was invented by a scam leader, and the 20+ members followed his template to scale up their business.

Based on a victim's self-account that was televised by China Central Television (CCTV) [3, 4], we summarize the setup, procedure and outcome of this cab sharing scam as follows.

Four scammers pretend not to know each other, but they work together. Each time, they work on a single victim (female).

First, the scammers identify a target (victim) in places where people would like to take taxi, e.g. a train station. A female scammer initiates a friendly conversation with the victim, who is waiting for taxi alone. They discover they are going to the same destination. Later, a male scammer comes by, and join their conversation. Three of them agree to share a cab.

A cab approaches them, with one male passenger on board already, and available seats are all in the back row. This is not surprising, since it is legitimate for four passengers to share a taxi, together with a driver.

Three passengers get into the back row, and the victim is seated in the middle.

The cab drives away, and then it stops at a destination. The front seat scammer rushes to get off the cab, intentionally dropping a wallet in the seat.

Another scammer grabs the wallet, shows off the money inside, and hides it for himself.

Soon, the departing scammer comes back to look for his wallet, claiming that two high-value bank cards are in the wallet, and they are all gone together. Nobody speaks up the wallet's whereabouts.

As confessed afterwards by herself, the victim initially hesitates to confront the scammer who has grabbed the wallet, worrying about his revenge e.g. being beaten up by him. Later, she attempts to speak up, but each time the female scammer has stopped her with hand gesture and persuasion like, "It's not your business, just stay away from it. It's fine to remain silent, if you're innocent and haven't taken the wallet".

The departing scammer becomes angry, and requests to search everyone's belongings (handbags). Nobody cooperates initially, but as tension builds up, the three scammers give up one by one.

The victim still refuses to cooperate.

The departing scammer changes his request, "I tell you how much each of the bank cards carry. You tell me your bank account, balance and PIN. With them we can check with your bank. If the account balance is the same as the amount you claim, then it's verified that the card is not the missing one, but indeed yours."

The victim finds it unacceptable to tell her card details to a total stranger. However, the two scammers, including the female one, does it without any hesitation. The female scammer starts to persuade the victim to follow suit, and so do the other scammers. Later, tension, pressure and suspicion on the victim is intensified.

With the desire of clearing herself of suspicion, the victim finally gives up. The front-seat scammer takes her card and phones her bank to check the account balance with her PIN. Then, he apologies to the victim and returns her the card, since she is proven innocent.

Next time the victim goes to ATM withdrawing money, but to no avail. Her card has been swapped by the front-seat scammer, and the money in her account are all gone.

### 2.3 Flight rebook scam

A victim receives a text message on his phone, a day or two before departure:

*"Mr Wang, due to mechanical failure (or weather problems), your flight CX234 from A to B scheduled for specific date/time is cancelled. Please phone our customer service number 400xxxx to change your flight or get refunded. Sorry for the inconvenience. CX Airline"*.

Everything in the message looks legitimate, and the victim makes a call to the customer service number.

When phone conversation starts, the victim is instructed to go to a nearby ATM machine.

At the ATM machine, the victim is asked to enter a number given by the scammer. It's an account to which money will go, but the victim does not realize it.

Once that number is entered, ATM indicates now to enter the amount to transfer.

The scammer continues his instruction to the victim, "Now, enters the first 6 digits of your card number for password verification."

Once the enter key is pressed, that amount gets transferred to the scammer's account, which will be emptied quickly.

The end of the story.

## 3 Why did the scams work?

We propose a simple theoretical framework to examine different factors that contribute to making the Chinese scams work. Then, we apply this framework to analyze each of the scams described above.

### 3.1 A simple framework

In our framework, three categories of factors all have a role in making a scam to work. First, *psychological* factors explain mind manipulation tricks that scammers apply to take advantage of victims. Second, *situational* factors are environmental or contextual settings that scammers build into their plot. For example, a cab was deployed in the cab-sharing scam to first create an isolated environment or context, and then to be filled with tension, and it constitutes an effective situational factor to make the scam work. Third, *social fabrics* are often exploited by scammers, too. For example, either weaknesses in socio-technical infrastructure or norms of a society scammers may take advantage of to fool victims.

Thus, our framework covers a wide spectrum of factors, ranging from individuals (humans), through contexts (environments) to a society at large (social fabrics).

We mainly use the well-known Cialdini's influence framework [5] to examine mind manipulations involved in a scam, and when necessary, we also refer to similar work by Stajano and Wilson [7], and by Lea et al [6]. The core of these three frameworks are more or less the same, as summarized by Stajano and Wilson themselves (see Table 1). However, the framework by Cialdini was the most original, and it has been tested both in research communities and in the field for more than 30 years.

Cialdini's framework is often summarized as "six weapons of influence", namely, reciprocity; commitment and consistency; social proof; authority; liking; and scarcity. Ross Anderson, a Cambridge academic, famously named these "a guide to pushing people's buttons".

### 3.2 Analyzing scams with our framework

We apply our theoretical framework to explain why each of the scams has worked.

**Criminal investigation scam.** We first look into psychological factors, then situational factors, and finally factors involving with social fabrics.

*Psychological.* The following methods were deployed for mind manipulation.

Appeal to authority. Impersonating a bank employee and impersonating a police officer both appeal to authority, aiming for the victim to take the conversation seriously and aiming for his or her obedience. Otherwise, it was so easy and effortless for the victim to quit by hanging the phone.

Intimidation. The victim was threatened by an immediate arrest and by ten years of imprisonment. These were both explicit intimidation. Appeal to fear with intimidation and appeal to authority are related concepts, but not the same. We will further clarify these in Section 4.

Scarcity. The scammers kept pushing the victim for reaction in a short amount of time. Once a shocking effect is created by intimidation, the scammers kept the victim's brain in a 'steamy' mode, giving her or him little chance to recover the brain from the shock.

*Situational.* Several situational factors were deployed to create a seemingly realistic scenario and context for a victim, e.g. emulating the standard greeting voice message used by genuine police stations; displaying a genuine police caller ID on a victim's phone. The legitimate-looking arrest warrant is a prominent and powerful situational factor deployed, and it helps to reinforce the intimidation effect and intensify the psychology of fear.

*Social fabrics.* This scam is a targeted attack, and the victim's personal information that the scammers used are all genuine, including name, national ID number, address, and phone number. Aggressive identify theft in one way or another is a weakness in social fabrics that this scam has exploited. These all help to make the accusation believable to at least such an extent that the victim feels, "Alas, I've had no wrongdoing, but I am in deep trouble now".

Moreover, sadly it is both a belief and practice that people can buy out of trouble or crime in some circumstances, if not always. This is another weakness in social fabrics that has played a role in making the scam work.

Some weaknesses in the banking system were exploited by the scammers, too. It is relatively easy for the scammers to open bank accounts with fake or stolen identities. There is little check and control for online money transfer, a scammer could easily empty a big sum from a victim account to his in one go, and he could easily withdraw a big sum from ATM machines in one day. On the other hand, the banks typically offer little help in tracing stolen money and recovering the money, since these are time-consuming but do not generate revenue for them.

**Cab sharing scam.** This is a well-engineered scam with multiple innovations.

*Psychological.* The scammers applied nearly all the weapons from Cialdini, except the authority method, for mind-manipulating the victim.

Reciprocity and liking: the female scammer is staged to apply these methods to get close with the victim, and to influence her behavior and decision making.

Social proof: all the scammers applied this method at least twice to creating a herding effect. The first use was to influence the victim to comply with the request of handbag search, and the second use was to let the victim share her bank account and PIN details.

Commitment and Consistency: The victim was hesitating whether she should speak up who took the missing wallet, and later when she wanted to speak up, she was stopped by the female scammer. This is an application of commitment and consistency, creating a psychological guilty effect on the victim, for she did not speak up the truth.

Scarcity: This method has been applied several times, and each time only a short while was allowed for the victim to make a decision and act accordingly.

In addition, an implicit but strong intimidation was applied to the victim, and this psychological pressure has been intensified by the choice of situational factors that will be discussed shortly.

*Situational.* The use of a moving cab, together with the intentional seat arrangement for the victim, i.e. the mid seat in the back row, created a strong sense of isolation, which leads to the feeling of helplessness and intimidation. According to her self-account, the victim says,

"I was entirely perplexed … I was feeling like in hot water; I did not know how to clear myself from something that I had not done; I was also nervous. In the meanwhile, the cab was still moving. I had a strong feeling of being isolated and helpless in such a narrow physical space. I was in a moving cab, so I was isolated from any external resources or people I could ask help from. I was also isolated from the surrounding people." [4]

*Social fabrics.* On the surface it appears no obvious weakness in social fabrics is exploited in this case. The truth is the opposite. The plot of taking a victim's bank card and PIN would not be possible, if a sloppy practice was not followed by China banks. All debit cards do not print an account holder's name, and this is a common practice in China. Therefore, the victim could not use her name printed on a bank card to argue that the card is hers.

In China, a savings account does not have a bank card, but goes with a hardcopy passbook where an account name is printed. On each credit card, an account name is printed, too. You cannot instantly get a credit card, since it takes time for a bank to double check income and credit history. The bank is also willing to bear with the hassle of printing an account name to please the customer – "This card is customised for you, my esteemed customer". However, a lot more debit cards are in circulation than credit cards. On the contrary, debit cards are mass produced without printing account names, and they are assigned to customers by an increasing or even random order of the card numbers. The banks have believed that it has no security consequences whatsoever by printing no account names on debit cards, and this has become a common practice in China for many years. The incentives for the banks are clearly cost-saving, operational efficiency and convenience to themselves.

However, the reality is that the scammers took advantage of the popular banking practice, and they had conveniently prepared fake debit cards for every major bank in their pockets.

Moreover, the banking flaws discussed for the criminal investigation scam are applicable here, too.

Another key design in this scam is that the scammers do not cross the line to turn it into a robbery. This way, they do not have to resort to brute force or to intimidate with weapons. Perhaps more importantly, scam often carries a much lighter punishment than robbery in China. Therefore, the scammers are more or less rational in this regard.

**Flight rebook scam.** On the surface, this scam looks extremely simple and hardly believable that it has made the top scam list. However, a strong psychological effect this scam creates by weaving multiple factors from all three dimensions.

*Psychological.* A 'time scarcity' effect is the only psychological factor embedded in this scam, since the legitimate-looking request comes near the last minute. The victim is given a short time span, and thus an urgency to rescue a relatively expensive ticket. However, many situational and social fabric factors are weaved into the scam, significantly reinforcing this 'time scarcity' effect.

*Situational.* The choice of near departure time is arguably a situational factor, and it is a necessary condition for the psychological factor of time scarcity to work.

The introduction of an ATM machine into the scene is a prominent situational factor in this scam. Victims are instructed to interact with a machine, rather than a friendly human whom you can talk to or even ask for help with. If they are unfamiliar with ATM use or its user interface sucks, there is really no choice better than following the instruction when they are being pushed by the scammer. In particular, during this part of interactions in the scam, the scammer intentionally gives the victim little time for thinking through and decision making.

The use of air tickets is another situational factor, and it works together with the second social fabric factor that is detailed shortly.

*Social fabrics.* First, flight delays and cancellations have been common in China. People have got used ('conditioned') to them and would rarely be suspicious about flight change requests. In the world's worst airport list, ranked by on-time-departure ratings in 2015, four Chinese airports made into top 5, including major air hubs in Hangzhou, Shanghai, Shenzhen and Nanjing [9]. A major reason behind the scene is the Chinese airspace is overcrowded and in short supply. Less than 30% of China's airspace is allocated for commercial airlines, compared to 80% of the airspace in the US for the same purpose [2]. On the other hand, the number of flights and air passengers keeps increasing, as China has risen to the second biggest economy.

Second, cheap airlines like EasyJet are rare in China, and thus travel by air is relatively expensive. Therefore, an air ticket is not something that everyone feels easy to dump, but something that deserves serious efforts to rescue.

Third, this scam is not a random shoot, but a targeted attack. All the flight and passenger information used in the scam is not fake, but genuine. Such sensitive infor-mation has been leaked in one way or another (e.g. via insider fraud or security flaws in computer systems), and available for sale in the underground market.

Finally, the banking flaws discussed for the criminal investigation scam are applicable here, too.

## 4   A Comparison with the Western scams

Three data sets for the common Western scams are available: 1) Phishing, Nigerian 419 scam & variants: these have been widely studied in the literature. 2) The scams from the BBC TV documentary "The Real Hustle" series by Paul Wilson, Alexis Conran and Jessica-Jane Clement. Over nine series, this team researched and documented hundreds of the scams most commonly carried out in Britain. This dataset includes various street scams, as well as modern ones like fake ATM in the street, PIN skimmer, and keyboard with key logging. 3) Various postal scams (studied in Lee et al [6]).

Our comparison of Chinese and western scams focuses only on psychology. We do not aim for a case-by-case comparison, which is tedious and not necessarily productive. Instead, we carry out a meta comparison.

The first set of scams, as researchers are familiar with, appeal to human beings' greedy. The second set of scams was carefully studied by Stajano and Wilson [7] to extract their framework for explaining human psychological vulnerabilities. The third set was carefully studied by Lee et al [6] to establish their own framework.

By combining all the principles established by the research on these three datasets with Cialdini's six weapons, we have a superset that covers all the principles established in the literature to explain the psychological factors exploited in the Western scams. With the aid of Table 1, it is easy to derive this superset, which ends up with ten elements, including reciprocity, commitment and consistency, social proof (or herding), authority (or social compliance), liking, scarcity, distraction, need and greed, kindness and dishonesty.

| Principle | Cialdini (1985–2009) | Lea et al. (2009) | Stajano-Wilson (2009) |
|---|---|---|---|
| Distraction | | ~ | ● |
| Social Compliance (a.k.a. "Authority") | ● | ○ | ○ |
| Herd (a.k.a. "Social Proof") | ● | | ○ |
| Dishonesty | | | ● |
| Kindness | ~ | | ● |
| Need and Greed (a.k.a. "Visceral Triggers") | ~ | ● | ○ |
| Scarcity (related to our "Time") | ● | ○ | ~ |
| Commitment and Consistency | ● | ○ | |
| Reciprocation | ● | | ~ |

● First identified this principle
○ Also lists this principle
~ Lists a related principle

**Principles to which victims respond, as identified by three sets of researchers.**

Table 1. A summary and comparison of the literature (taken from [7])

However, this superset seems to explain psychological tricks used in the Chinese scams only to an extent. As an additional mechanism, intimidation (which works by creating fear) is prominently spelled out in the Chinese scams, but not in the western counterparts (or at least not prominent enough to motivate the researchers to single it out as a separate mechanism or principle).

Appeal to authority (or social compliance) and intimidation are related concepts, but they are not the same. For example, a common interpretation of Cialdini's authority notion is the following. "Authority – People will tend to obey authority figures, even if they are asked to perform objectionable acts. Cialdini cites incidents such as the Milgram experiments in the early 1960s and the My Lai massacre."

Stajano and Wilson's description of their social compliance principle goes as follows. "Society trains people not to question authority. Hustlers exploit this "suspension of suspiciousness" to make you do what they want."

The jewelry-shop scam used by Stajano and Wilson to highlight their social compliance principle [7] works as follows. A scammer attempts to buy an expensive jewelry, and two accomplices come to 'arrest' her by impersonating plainclothes police officers. The 'cops' expose her as a notorious fraudster known for paying with counterfeit cash, collect as evidence the cash she's using as well as and the jewelry, which the 'cops' say 'will be returned', and then leave. The jeweler believes the jewelry will be returned, and she is grateful that the police saves her from the fraud. Appeal to authority explains this scam, but intimidation is clearly not part of it.

On the other hand, in the cab-sharing scam, intimidation employed there is not about authority at all.

In a nutshell, intimidation directly appeals to (the psychology of) fear, and it is a concept related to, but not the same as, appeal to authority (or, social compliance).

The psychology of persuasion was initially proposed by Cialdini for marketing professionals, and it is hardly effective to sell goods or services via intimidation, in particular for long term. Therefore, the omission of 'appeal to fear' in his framework was natural. On the contrary, ingenious Chinese scammers do not mind at all deploying intimidation, a coercive and powerful mind-manipulating mechanism, to pull off their scams.

## 5    Countermeasures

The human vulnerabilities identified in prior art [5, 6, 7] all help to understand scams. In particular, Stajano and Wilson [7] offered an insight that security engineers should design systems that are robust to such human vulnerability. When convenient, they also discussed how to apply some principles to secure computer systems.

We believe our PSS framework provides a new and systematic guidance for exploring scam countermeasures. First, it provides a multi-facet perspective to examine scams and countermeasures, and it cover people, scam context, and social fabrics.

Second, our framework suggests the following insights. While we should and need take into consideration of human vulnerabilities in designing secure systems, the cab-sharing scam, perhaps as an extreme case, clearly suggests that we cannot 'fix' human beings to defend against scams. It is sometimes simply beyond one person's capability to protect herself from scams. The dimension of situational factors is not the place we should aim to build defense in, either. Scammers are inventive, and it is hardly possible for defenders to either predict or address all situational factors that the scammers may make use of. Instead, the dimension of social fabrics is the right place to defend against scams. We should and need identify weaknesses in our socio-technical infrastructure that scammers may exploit, fix these vulnerabilities, and improve our social fabrics to maintain a healthy society that is robust and resilient to various scams, including both the known ones and those to be invented.

Intuitively, these observations make sense, too. In a complex social-technical system such as a rapidly developing society, if on our own, we need much more knowledge and capability than what we might realize or expect, in order to protect ourselves. Instead, individuals should rely on the dependability of complex socio-technical systems. That is, a robust and resilient socio-technical system should be in place in a society to protect their citizens.

With this fresh angle, we explore countermeasures for the Chinese scams and have the following observations. First, banks are behind the scene facilitating the flight rebook scam, the cab-sharing scam and the criminal investigation crime; they are beneficiary of all the three scams, too. Some effective scam countermeasures are really about improving banking security, for example: 1) Setting up a daily upper limit for ATM money withdrawal limit -- reportedly, many big sums were transferred to the scammers' accounts in one go, without difficulty or any question asked. In the USA, anti-money laundering laws and regulations at the federal and state levels require tracking and reporting each transaction of USD 5,000 or higher. However, similar mechanisms appear to be not in place in China, or not enforced strictly. 2) Alerting or disallowing a large-amount transfer from a bank account to a stranger (or unregistered) account.

These banking security mechanisms are hardly novel or difficult to implement, but Chinese banks are geared towards efficiency and convenience, instead. Presumably, the Chinese banking system did not take into consideration human vulnerabilities in their security design at all.

On the other hand, telecom companies were behind the scene facilitating the flight rebook scam and the criminal investigation crime; they are beneficiary of both scams, too.

According to a respected insider, banking and telecom industries profit from a significant cut, about 13%, of the scam value chain in China; they had done little to counter the rising scam threats, and they should and could do a lot more [8]. This suggests further weaknesses in social fabrics: misaligned incentives and a failed placement of liability.

It will help with scam countermeasures to fix these problems, and these fixes apparently should be carried out in the dimension of social fabrics, rather than the dimensions of psychological or situational factors.

One more point we want to make here: a reasonable method for designing solutions to address human vulnerabilities appears to examine scams in their entire socio-technical setting (i.e., treating the society as a large socio-technical system).

It is hardly possible to prevent victims from suffering from scams, but the right countermeasure should rely on improving the socio-technical system of our society to protect people (and their assets). Our PSS framework offers an analytical tool that does more or less an architecture analysis for resolving socio-technical security challenges. Little can be done with individuals and situational factors in terms of scam countermeasures, but much can and should be done in the dimension of social fabrics.

## 6  Conclusions

We have introduced and examined three high-profile Chinese scams, namely the criminal investigating scam, the flight rebook scam and the cab sharing scam. Not necessarily every element in the scams is novel. However, new twists are introduced to blend with known elements, and they together make some novel genres.

These scams have some distinctive features. First, Chinese scammers adapt their schemes well to the cultural and social context. We doubt these scams would work in the western countries like Britain. For example, the criminal investigating scam would only work in authoritative countries. The flight rebook scam would rarely work in a place where flight delays and cancellations do not occur (often). On the contrary, some western scams (like phishing) happen in China, and they work effectively even if they are pulled off verbatim. Second, these scams were sophisticated and well-engineered; well-organized with a division of labor and teamwork; and often with scripts available for internal training. In particular, the cab sharing scam is one of the best engineered scams we have seen so far. Third, these scams have been surprisingly lucrative.

We have proposed a simple but novel framework to understand why these scams have worked. Our PSS framework examines psychological, situational and social fabric factors that have played a role in each scam. A key observation is that the scammers hacked not only human psychology and people's unpreparedness – these scams were novel to many people and they were caught off guard – but also systematic weaknesses in the rapid changing society driven by the fast-growing economy, such as a banking system geared towards efficiency and convenience, failed placement of liability and incentives, and lack of protection for personal identify information and sensitive business transactions.

As a side note, the well-known Cialdini's influence framework [5], or the similar work by Stajano and Wilson [7] or by Lea et al [6], explains the psychological tricks used in the Chinese scams, but only to an extent. Revisions and extensions to the established scam or persuasion principles seem necessary to account for ingenious scammers in China. For example, the Chinese scammers do not mind resorting to intimidation. Either explicit intimidation (as in the criminal investigation scam) or implicit intimidation (as in the cab sharing scam) is evident.

As an attempt of looking for general countermeasures, we have argued that instead of fixing human beings or scrutinizing situational factors, it is more realistic and promising to address these scams by fixing the weaknesses in the socio-technical infrastructure and improving social fabrics.

## 7  Acknowledgement

This work was presented at "*Decepticon: 2nd Int'l Conference on Deceptive Behavior*" (Stanford, August 2017) and at the "*Security and Human Behaviour workshop*" (Carnegie Mellon, May 2018). We were grateful for valuable feedbacks and comments received at both events.